\documentstyle[12pt,epsfig,amssymb]{article}
\begin{document}
\setlength{\parskip}{0.45cm}
\setlength{\baselineskip}{0.75cm}
%
\begin{titlepage}
\begin{flushright}
May 2001 \\
\end{flushright}
\vspace*{1.2cm}
\begin{center}
\Large
{\bf {Impact of Higher Order and Soft Gluon Corrections}}

\vspace{0.1cm}
{\bf {on the Extraction of Higher Twist Effects in DIS}}

\vspace*{1.6cm}
\large
{S.\ Schaefer$^a$, A.\ Sch\"afer$^a$, M.\ Stratmann$^{a,b}$}

\vspace{0.8cm}
\normalsize
{$^a$ Institut f\"ur Theoretische Physik, Universit\"at Regensburg,\\
D-93040 Regensburg, Germany}

\vspace{0.1cm}
{$^b$ C.N.\ Yang Institute for Theoretical Physics, SUNY at Stony Brook,\\ 
Stony Brook, NY 11794-3840, USA}

\vspace*{1.6cm}
\large
{\bf Abstract}\\
\end{center}
\vspace{0.2cm}
The impact of recently calculated next-to-next-to-leading order QCD corrections 
and soft gluon resummations on the extraction of higher twist contributions to
the deep-inelastic structure function $F_2$ is studied using the BCDMS and
SLAC data.
It is demonstrated to which extent the need for higher twist terms is diminishing
due to these higher order effects in the kinematical 
region, $0.35\le x\le 0.85$ and $Q^2>1.2\;\mathrm{GeV}^2$, investigated. 
In addition, theoretical uncertainties in the extraction of higher twist contributions
are discussed, and comparisons to results obtained previously are made.
\end{titlepage}
\newpage
%
Deep-inelastic lepton-hadron scattering (DIS) data are an integral part
of all global QCD analyses of parton distribution functions \cite{ref:mrst,ref:pdf}.
To guarantee that only the leading-twist part of the parton densities is
extracted usually only data with sufficiently high momentum transfer $Q^2$
and invariant mass $W^2=Q^2(1/x-1)$ are selected, e.g., $Q^2>2\;\mathrm{GeV}^2$ and
$W^2>10\;\mathrm{GeV}^2$ \cite{ref:mrst}.
These universal distributions can then be applied to other hard 
processes by virtue of the factorization theorem.
Contributions of higher twist (HT) are expected to become increasingly important
as $x\rightarrow 1$ as shown, e.g., in the infrared renormalon approach \cite{ref:beneke},
but they are suppressed by additional powers of $1/Q^2$
with respect to the logarithmic $Q^2$ dependence of the leading twist contribution.
Hence HT operators can be safely neglected in all conventional global QCD analyses 
\cite{ref:mrst,ref:pdf} due to the aforementioned cuts on $Q^2$ and, in particular, on $W^2$.

However, twist-four contributions are interesting in themselves and 
can provide valuable insight into higher quark-quark and quark-gluon correlators
inside the nucleon. Not very much is presently know about these correlators
apart from some studies in the framework of lattice QCD \cite{ref:lattice}.
Recent measurements of the transversely polarized DIS structure function
$g_2$ \cite{ref:g2data} may indicate that at least some HT are small\footnote{Measurements of 
$g_2$ provide a unique possibility to disentangle its twist-three part
since the leading twist part can be entirely expressed in terms of the 
rather well known structure function $g_1$.} for an averaged
$Q^2$ of about $2\div3\;\mathrm{GeV}^2$ even at fairly large values of $x$.
A better understanding of HT contributions, in particular their importance in
describing low $Q^2$ DIS data, is important in many respects. A wealth of 
low $Q^2$ DIS data presently discarded in global QCD analyses would open up.
Secondly, future experiments, e.g., the `CEBAF @ 12 GeV' program, focus on measurements
at high $x$ at comparatively low $Q^2$ where HT are expected to be relevant. 
Finally, QCD analyses of longitudinally polarized
DIS data would benefit. Here the available data are too scarce to
allow for sufficiently `safe' cuts on $Q^2$ and $W^2$ when 
extracting leading twist polarized parton densities,
and the importance of HT in relating the measured spin asymmetry
$A_1 \simeq g_1/F_1$ (where HT may partly cancel in the ratio) to the
structure function $g_1$ is still an open issue, see, e.g.\  Ref.\ \cite{ref:grsv}.

At present, data on the unpolarized structure function $F_2$ obtained 
quite some time ago by the CERN-BCDMS and various SLAC experiments \cite{ref:data} 
still offer the best testing ground for HT studies since they cover a wide kinematical 
range up to large $x$ and down to low values of $Q^2$ with sufficiently good 
statistical accuracy. Therefore several attempts have been already made to disentangle
leading and higher twist contributions to $F_2$ [8-14].
As anticipated, HT effects become increasingly important as $x\rightarrow 1$.
Based on a partial next-to-next-to-leading order (NNLO) QCD analysis 
it was argued \cite{ref:bodek2} that the need for HT is diminishing greatly 
when going from the NLO to the NNLO of QCD\footnote{It
should be noted that similar studies have been made also for the charged
current structure function $xF_3$ \cite{ref:f31,ref:f32}. In \cite{ref:f32} 
it was also observed that NNLO contributions largely remove the need for 
HT at large $x$.}. A first complete NNLO analysis \cite{ref:mrstht2}
could not confirm this observation. However, in that paper no distinction between
different sources of HT terms was made, and it could be that most of the remaining HT
in NNLO is of purely kinematical origin and hence calculable (see discussion below
Fig.\ 1).

In the remainder of the paper we further extend these studies 
making use of the NNLO coefficient functions \cite{ref:nnlocoeff} and
recent estimates for the full $x$ dependence of the
NNLO splitting functions \cite{ref:vogt} based on the integer Mellin $n$
moments calculated in \cite{ref:moments}.
Another important ingredient of our analyses will be soft gluon resummations (SG) 
which have not been considered so far in extractions of HT contributions 
to DIS. The quark coefficient functions $C_{1,2}^q$ 
in DIS which link the quark distributions to 
the structure functions $F_{1,2}$ exhibit a large $x$ enhancement of the form
$[\ln ^{l-1} (1-x)/(1-x)]_+$ where $l=1,\ldots,2k$ in the ${\cal{O}}(\alpha_s^k)$
approximation to $C_{1,2}^q$ which needs to be resummed 
to all orders \cite{ref:resum,ref:vogtsg}. 
Finally, we compare the outcome of our analyses with results already available in the
literature.
It should be stressed that we mainly focus on a qualitative comparison of HT
contributions to $F_2$ extracted in NLO, NNLO, and NNLO including SG and do not
attempt to provide a full set of parton densities or to extract $\alpha_s(M_Z)$
from BCDMS and SLAC $F_2$ data as was done, e.g., in \cite{ref:mv,ref:alekhin}. 
Instead we study the various sources of theoretical uncertainties
in the determination of HT contributions to DIS.

%
%
The DIS structure function $F_2$ can be expanded in $1/Q^2$ as
\begin{equation}
\label{eq:twist}
F_2(x,Q^2) = F_2^{(2)}(x,Q^2) +  F_2^{(4)}(x,Q^2)/Q^2 + {\cal O}(1/Q^4)
\end{equation}
where $F_2^{(t)}$ denotes the contribution of twist-$t$.
The leading twist part $F_2^{(2)}$ `factorizes' into a convolution of 
a perturbatively calculable coefficient function $C_2$ and some non-perturbative
parton density combination $f$ which can only be determined from experiment so far:
\begin{equation}
\label{eq:f2}
F_2^{(2)}(x,Q^2) = x [C_2(\alpha_s(\mu_r^2), \frac{Q^2}{\mu_f^2}, \frac{\mu_f^2}{\mu_r^2}) 
\otimes f(\mu_f^2,\mu_r^2)](x)\;\;\;.
\end{equation}
Once the non-perturbative input $f$ is fixed at some reference scale
$\mu_0$ its $\mu_f$ dependence is fully predicted by the well-known
DGLAP evolution equation which schematically reads
\begin{equation}
\label{eq:pdfevol}
\frac{\mathrm{d}}{\mathrm{d} \ln \mu_f^2} f(x,\mu_f^2,\mu_r^2) = 
[{\cal P}(\alpha_s(\mu_r^2),\frac{\mu_f^2}{\mu_r^2})
\otimes f(\mu_f^2,\mu_r^2) ](x) \;\;\;,
\end{equation}
where $\mu_f$ and $\mu_r$ denote the factorization and renormalization
scales, respectively.
For simplicity we have limited ourselves in 
Eqs.\ (\ref{eq:f2}) and (\ref{eq:pdfevol}) to the non-singlet sector 
which is all what is needed for our analyses as will be discussed below.

The coefficient function $C_2$ in (\ref{eq:f2}) is known up to NNLO \cite{ref:nnlocoeff}
while only the first integer moments of the evolution kernels ${\cal P}$ in three-loop order
have been calculated so far \cite{ref:moments}. However, based on these
results and further constraints on ${\cal P}$  estimates for the full
$x$ dependence of the NNLO kernels have been derived recently \cite{ref:vogt}.
The residual uncertainties on the kernels were shown to be extremely small \cite{ref:vogt}
in the large $x$ region, $x\gtrsim 0.3$, we are interested in.
The relevant coefficients of the QCD beta function up to $\beta_2$ which govern the 
running of $\alpha_s(\mu_r)$ are given in \cite{ref:qcdbeta}.
The appropriate matching conditions at flavor thresholds can be found in \cite{ref:matching}.
We are using the $\overline{\mathrm{MS}}$ scheme throughout this paper.

There are two sources of power corrections in $1/Q^2$ contributing to
$F_2(x,Q^2)$ in (\ref{eq:twist}) beyond twist-2.
The first one is of purely kinematical origin and can be entirely expressed 
in terms of the leading twist contribution $F_2^{(2)}(x,Q^2)$ in (\ref{eq:f2}).
It only takes into account effects of the so far neglected non-zero target mass $M$ and
approximately behaves like $x^2 M^2/Q^2$. Hence it gives a
sizable contribution to $F_2$ at large $x$ whenever $Q^2$ is of ${\cal{O}}(M^2)$ or smaller.
The `target mass corrected' (TM) expression for $F_2^{(2)}(x,Q^2)$ is know for a long
time and reads \cite{ref:gp}
\begin{eqnarray}
\nonumber
F_2^{\mathrm TM}(x,Q^2) &=& \frac{x^2/\xi^2}{r^{3/2}} F_2^{(2)}(\xi,Q^2)
+6 \frac{ M^2}{Q^2} \frac{x^3}{r^2} \int_\xi^1
\frac{\mathrm{d} \xi'}{{\xi'}^2} F_2^{(2)}(\xi',Q^2) \\
&&
\label{eq:tmc}
+12 \frac{ M^4}{Q^4} \frac{x^4}{r^{5/2}} \int_\xi^1 {\mathrm{d}} \xi' 
\int_{\xi'}^1 \frac{\mathrm{d} \xi''}{{\xi''}^2} F_2^{(2)}( \xi'',Q^2)\;\;\;,
\end{eqnarray}
with $\xi = 2x/(1+\sqrt{r})$ and $r = 1+4 x^2 M^2 /Q^2$.
It is sometimes preferred to use $F_2^{\mathrm TM}$ only after expanding
(\ref{eq:tmc}) in powers of $M^2/Q^2$ and retaining only the leading term.
We will use both, the full and the expanded expressions for
$F_2^{\mathrm TM}$ in our fits. Any differences somehow represent part of
the theoretical uncertainty in the extraction of higher twists from
DIS data.

The other source of power corrections cannot be related to $F_2^{(2)}(x,Q^2)$ in general
and provides important new insight into the QCD dynamics of
higher quark-quark and quark-gluon correlations in 
the nucleon about which almost nothing is known yet. Also the $Q^2$ evolution of these 
twist four operators has not been calculated yet\footnote{\label{foot:htevol} Even if the evolution
of twist four operators would become available, it would be extremely difficult to make 
use of it because the twist-4 operators which are accessible in DIS mix under evolution with other
twist-4 operators which do not contribute to DIS and have to be taken from elsewhere.}.
Therefore one has to fully rely on some model here. One interesting possibility
is the infrared renormalon approach \cite{ref:beneke} which allows to calculate explicitly
the power corrections to the coefficient function $C_2$ in (\ref{eq:f2}), i.e., to
predict the $x$ shape of $F_2^{(4)}(x,Q^2)$ up to some unknown normalization factor,
which has to be determined from data, see \cite{ref:webber} for details.
This approach was used in some of the previous analyses of higher twists in
DIS \cite{ref:alekhin,ref:bodek1,ref:bodek2}. It turned out that, although the renormalon
model could reproduce the general trend that HT increase as $x\rightarrow 1$, the 
predicted $x$ shape was somewhat off (see, e.g., Fig.\ 2 in \cite{ref:alekhin}).
Therefore we prefer not to fix the $x$ shape of $F_2^{(4)}(x,Q^2)$ and instead use
a simple multiplicative ansatz\footnote{\label{foot:ansatz}Sometimes $F_2^{(4)}(x,Q^2)$ is defined
in such a way that it does not exhibit any $Q^2$ dependence, i.e.,
$F_2(x,Q^2)=F_2^{\mathrm{TM}}(x,Q^2)+ \tilde{C}_{\mathrm{HT}}(x)/Q^2$ 
\cite{ref:alekhin2}.} as was also employed in the first HT 
analysis by Milsztajn and Virchaux \cite{ref:mv} as well as in \cite{ref:bodek1,ref:bodek2}.
For each $x$ bin of the data we simply introduce a factor $C_{\mathrm{HT}}$ which is
independent of $Q^2$, 
\begin{equation}
\label{eq:htdef}
F_2(x,Q^2) = F_2^{\mathrm{TM}}(x,Q^2) \left[ 1+ \frac{C_{\mathrm{HT}}(x)}{Q^2}
\right] \;\;\;,
\end{equation}
i.e., $F_2^{(4)}(x,Q^2)$ is {\em approximated} to have the same logarithmic 
dependence on $Q^2$ as $F_2^{(2)}(x,Q^2)$. 

Finally, we also study the influence of soft gluon resummations for the coefficient
function $C_2$ in (\ref{eq:f2}) \cite{ref:resum,ref:vogtsg}
on the extraction of HT from DIS data. Since these resummations 
are operative also in the large $x$ region where HT become important
they may have a sizable impact on the size of the 
extracted HT coefficients $C_{\mathrm{HT}}$ in (\ref{eq:htdef}).
The SG take care of potentially large logarithms of the form
$[\ln ^{l-1} (1-x)/(1-x)]_+$ where $l=1,\ldots,2k$ in the ${\cal{O}}(\alpha_s^k)$
approximation to $C_2$ which threaten to spoil the convergence of the
perturbative expansion by resumming them to all orders in $\alpha_s$.
These resummations have been recently pushed up to 
next-to-next-to-leading logarithmic terms \cite{ref:vogtsg} and 
can be straightforwardly implemented in the Mellin $n$ moment space
which we also use for solving the evolution equations (\ref{eq:pdfevol}).

Before we turn to our numerical results let us specify the 
ansatz for the parton density function $f$ in 
(\ref{eq:f2}) and the data sets and cuts used in our analyses.
Since we are only interested in the impact of NLO, NNLO, and SG
corrections on the size of the higher twist coefficients
$C_{\mathrm{HT}}$ in (\ref{eq:htdef})
we can limit ourselves to the large $x$ region by taking into
account only the BCDMS and SLAC data (using the BCDMS $x$ binning) with $x\ge 0.35$
\cite{ref:data}. This is also the region
where the estimates for the NNLO kernels ${\cal P}$ work extremely
well \cite{ref:vogt}.
As in [10-13] 
we then apply the non-singlet (`valence') approximation
for $F_2$ and only a single combination of parton densities $f$ is
required for proton target data\footnote{We refrain from using 
deuterium data \cite{ref:data} as well. They are equally precise than proton
data but may suffer from nuclear binding effects.
We also discard SLAC data with $x>0.85$ \cite{ref:slac2} which are
close to or in the resonance region.}
(we will comment below on a possible `contamination' of the scaling 
violations of $F_2$ at large $x$ due to singlet and
gluon contributions). In addition we select only data with
$Q^2\ge 1.2\,\mathrm{GeV}^2$ to stay away from the resonance region 
and the region where the perturbative expansion is expected to break down since
$\alpha_s$ becomes too large. Six bins in $x$ remain after our cuts, so we
have to determine six different HT parameters
$C_{\mathrm{HT}}(x=0.35),\ldots, C_{\mathrm{HT}}(x=0.85)$ in (\ref{eq:htdef}) 
from the fit to the data.

We parametrize the input distribution $f(x,\mu_0^2)$ in (\ref{eq:f2})
by the standard ansatz
\begin{equation}
\label{eq:ansatz}
f(x,\mu_0^2) = N x^\alpha (1-x)^\beta (1+ \gamma_1 \sqrt{x}+\gamma_2 x) \;\;
\end{equation}
using $\mu_0 = 1\,\mathrm{GeV}$ as the initial scale. It turns out, however, that all
fits are rather insensitive to $\gamma_1$ and $\gamma_2$, so we will 
neglect these terms in the following.
$\alpha_s(\mu_r)$ is always computed by solving its renormalization
group equation exactly in NLO or NNLO since the approximate formula \cite{ref:pdg}
is not sufficient for small scales.
For simplicity we always choose $\mu_r=\mu_f=Q$.
The evolution equations (\ref{eq:pdfevol}) are most conveniently
solved in Mellin $n$ space. There are, however, different ways how to
actually truncate the solution at a given order $k$ in perturbation theory
($k=1$: NLO, $k=2$: NNLO), see, e.g., \cite{ref:evolsolve} for a
detailed discussion. On the one hand one can have a sort of
iterative solution where all orders in $\alpha_s$ still contribute,
on the other hand one can impose strict power counting by keeping
only terms up to ${\cal{O}}(\alpha_s^k)$. Both approaches are widely
in use in global QCD analyses and differ, of course, 
only by terms of  ${\cal{O}}(\alpha_s^{(k+1)})$.
This `freedom' also represents part of the theoretical uncertainty and we
will perform our fits using different solutions to (\ref{eq:pdfevol}).

We should also mention that we have always added 
the statistical and systematic errors in quadrature 
as is commonly done in most QCD analyses of parton densities.
In \cite{ref:mv} {\em some} of the systematic errors of the BCDMS data
have been combined to a so called `main systematic error' \cite{ref:shift}.
The BCDMS data points are then allowed to float within this error with
a common normalization factor to be determined from the $\chi^2$ fit \cite{ref:mv}.
The BCDMS data shifted in this way are used in some of the global QCD
analyses. Unless stated otherwise, we 
refrain from using these data \cite{ref:mv} as they 
are strongly biased by the theoretical input that went into the analysis,
and we stick to the original data sets \cite{ref:data} instead.
In particular, it should be noticed that in \cite{ref:mv} data with
$Q^2$ values as low as $0.5\,\mathrm{GeV}^2$ were taken into account and
that only the approximate solution for the running of $\alpha_s$ \cite{ref:pdg} was used
which is known to be way off the exact solution at such small scales.
In addition, it has been shown recently in \cite{ref:alekhin} that
the introduction of the main systematic error is not a rigorous method
to deal with systematic errors and that correlations should be
fully taken into account to obtain reliable errors for the
parameters extracted from a $\chi^2$ analysis.
Nevertheless, as we are mainly interested in the qualitative effects
of higher order corrections we treat all errors as uncorrelated
which greatly simplifies the analysis and should not effect the
general features of our results. The statistical meaning of the
$\chi^2$ function is totally obscured in QCD analyses of parton densities
anyway, and errors are very difficult to quantify \cite{ref:tung}.
Therefore the errors for the $C_{\mathrm{HT}}$ as obtained in our
$\chi^2$ fits should not be taken too literally or 
be regarded as $1\sigma$-errors.
As in \cite{ref:mv,ref:bodek1,ref:bodek2} we allowed for a global normalization shift 
of the BCDMS data with respect to the SLAC data.  

\begin{figure}[tb]
\begin{center}
\epsfig{file=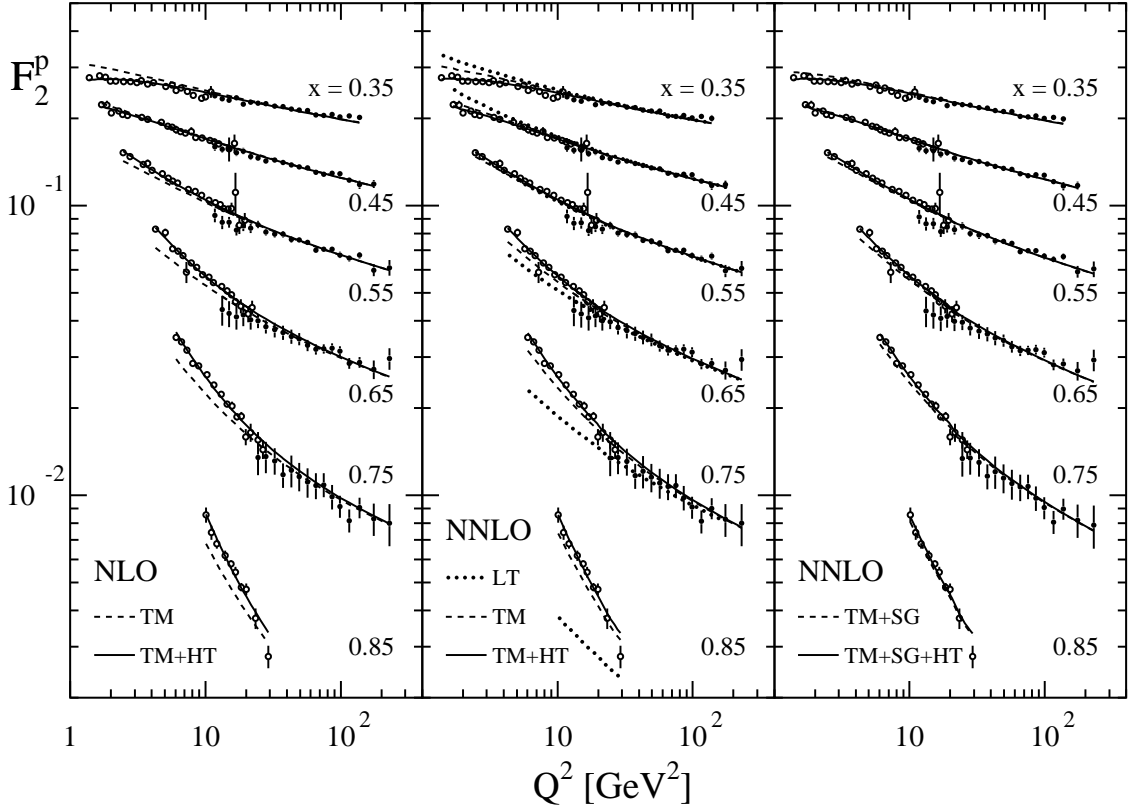,width=0.97\textwidth}
\vspace*{-0.2cm}
\caption{\sf Comparison of the results of our $\chi^2$ fits (solid lines) 
in NLO and NNLO of QCD and in NNLO including SG 
with the SLAC and BCDMS proton target data \cite{ref:data} (open and full circles, 
respectively) using $\alpha_s(M_Z)=0.117$.
The dashed lines refer to the target mass corrected part
$F_2^{\mathrm{TM}}(x,Q^2)$ only with the additional higher twist
factor $(1+C_{HT}(x)/Q^2)$ in Eq.\ (\ref{eq:htdef}) being omitted.
The leading twist (LT) part $F_2^{(2)}$ without TM is shown as well in the center
part (dotted lines) for illustration.}
\end{center}
\end{figure}
In Fig.\ 1 we compare the results of our $\chi^2$ analyses of $F_2$
according to Eq.\ (\ref{eq:htdef}) (solid lines) with the
SLAC and BCDMS proton target data \cite{ref:data} using
$\alpha_s(M_Z)=0.117$. Very good agreement 
with very  similar values of $\chi^2$ is achieved
for the fits in NLO and NNLO QCD as well as for a fit in NNLO including the
SG mentioned above.
The size of the required dynamical HT contributions can be inferred from comparing the
dashed lines, which refer to the TM corrected part of $F_2$ only 
with $(1+C_{\mathrm{HT}}(x)/Q^2)$ in Eq.\ (\ref{eq:htdef}) being omitted,
with the full results (solid lines). Clearly the need for HT contributions
is diminishing when going from NLO to NNLO. A further reduction, 
most noticeable at $x=0.85$, is observed when the SG are taken into account as well. 
In the center part of Fig.\ 1 we also give the
leading twist (LT) result $F_2^{(2)}$ without target mass correction. A comparison with the 
corresponding TM and full results shows that a major part of the HT contributions 
required to describe the data is of purely kinematical rather than dynamical origin
and hence calculable.
The extracted higher twist coefficients $C_{\mathrm{HT}}$ for the six 
$x$ bins are shown separately on the left hand side (l.h.s.) of Fig.\ 2.

\begin{figure}[tb]
\begin{center}
\epsfig{file=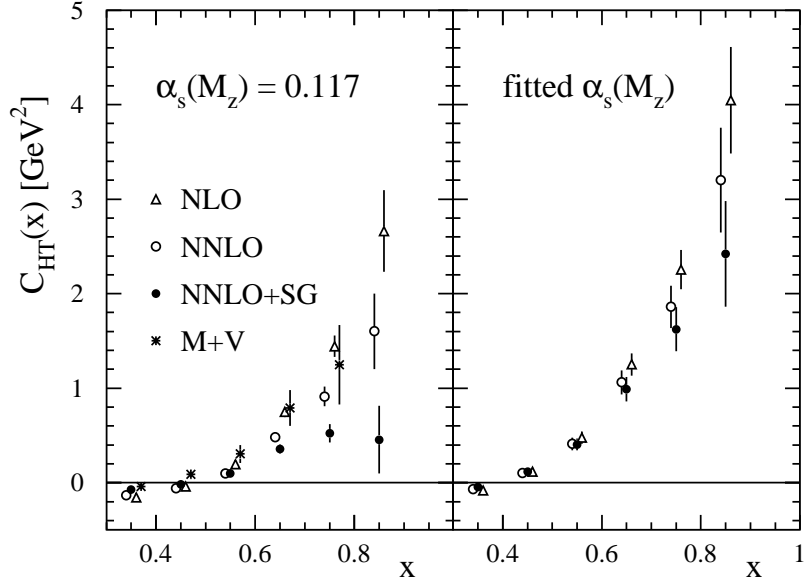,width=0.7\textwidth}
\vspace*{-0.8cm}
\caption{\sf Extracted higher twist parameters $C_{\mathrm{HT}}$ in six
$x$-bins according to Eq.\ (\ref{eq:htdef}) 
for fixed and fitted values of $\alpha_s(M_Z)$. 
On the l.h.s.\ also the results obtained
in \cite{ref:mv} are shown for comparison (crosses).
For each $x$-bin the results are slightly horizontally
displaced to avoid overlapping error bars.}
\end{center}
\end{figure}
It should be mentioned that we have also tried to obtain $\alpha_s(M_Z)$ from
the fits rather than setting it to a fixed value. 
Unfortunately, the values we obtained turned out to be unacceptably 
small, $\alpha_s(M_Z)\approx 0.105$, way outside the current world
average even within the error bars.
The gross feature that the HT parameters get smaller when going to 
higher orders of perturbation theory is common to all results
but less pronounced for $\alpha_s(M_Z)\approx 0.105$ as can be seen
by comparing the left and the right hand sides of Fig.\ 2.
This also demonstrates the strong correlation between the size
of the $C_{\mathrm{HT}}$  coefficients and the value of $\alpha_s(M_Z)$:
the smaller the $\alpha_s(M_Z)$ the larger the HT contribution as was
already pointed out in \cite{ref:alekhin}.

We have investigated the reason for not obtaining a more common
value for $\alpha_s(M_Z)$ from our fits in particular because
reasonable values for $\alpha_s(M_Z)$ have been obtained in previous
analyses \cite{ref:mv,ref:alekhin}. The latter fits differ, however, in
various aspects from our ones. In \cite{ref:mv} a NLO fit to all
available BCDMS and SLAC proton and deuterium data was 
performed\footnote{\label{foot:mv} As mentioned above, the approximate expression for 
$\alpha_s(\mu_r)$ \cite{ref:pdg} was used in \cite{ref:mv} which is not valid for small
scales. In addition not all details of their analysis are
specified in \cite{ref:mv}. Therefore it is difficult to actually compare the result for
$\alpha_s(M_Z)$ (and the HT coefficients) obtained in \cite{ref:mv} 
with our NLO results on a quantitative level.}
in the range $Q^2>0.5\,\mathrm{GeV}^2$ and $0.07\le x \le 0.75$.
If we extend the range of data included in our fit to $0.07\le x$,
but still requiring $Q^2>1.2\,\mathrm{GeV}^2$, and introduce also a singlet and
gluon contribution to $F_2$, we obtain much more reasonable values for 
$\alpha_s(M_Z)$ in NLO as well.
This indicates that $\alpha_s$ cannot be extracted from large $x$
data alone and/or that singlet and gluon distributions are still important when
studying scaling violations of $F_2$ for $x>0.35$, i.e., that
a pure non-singlet approximation for $F_2$ is not sufficient here.
Another possible explanation for the small value of $\alpha_s(M_Z)$ may come
from our simplified treatment of the systematical errors.
To investigate this further we have performed similar fits using the data
in \cite{ref:mv} which include the BCDMS main systematic error shift mentioned above.
In NLO this also leads to a significantly increased value for 
$\alpha_s(M_Z)$ but almost no changes occur in NNLO.
This does not come as a surprise since the main systematic error shift in \cite{ref:mv}
was obtained only from a NLO fit, and it strongly depends on the theoretical
input. It seems to be reasonable to assume that
an $\alpha_s(M_Z)$ of about $0.116\div 0.118$ would  be obtained 
in our fits once the correlations between different sources of systematical
errors are properly taken into account as in \cite{ref:alekhin,ref:alekhin2}. 
This is however far beyond the scope of our more qualitative studies.
Data at smaller values of $x$ may also be important but a reliable NNLO analysis
can only be performed once the complete NNLO kernels become available.
As in \cite{ref:bodek1,ref:bodek2}, we therefore prefer 
to fix $\alpha_s(M_Z)$ in all our fits to 0.117 rather than
taking the fitted value which just seems to be an artifact of the
approximations used in our analyses.

Before we turn to a comparison with results obtained previously in the literature, 
let us briefly focus on the theoretical uncertainties in the extraction of $C_{\mathrm{HT}}$
[and $\alpha_s(M_Z)$] from DIS data. We have already mentioned that there
are different ways of how to treat the TM in (\ref{eq:tmc}) and that there
is no unique way of
actually solving the evolution equations (\ref{eq:pdfevol}) to a given order.
We have performed several fits to take this ambiguity into account and it turns out that
the gross features of $C_{\mathrm{HT}}(x)$ are basically unchanged but the
precise values of the $C_{\mathrm{HT}}(x)$ do depend to a certain extent on
the details of the analysis.
For simplicity we have always chosen $\mu_r=\mu_f=Q$ but nothing prevents
us from varying these two scale independently around $Q$ which would
also alter the values of the $C_{\mathrm{HT}}(x)$.
The dependence on the factorization scheme and on $\mu_f$ can be removed if one
expresses the scaling violations of $F_2$ in terms of $F_2$ itself
using `physical' evolution kernels, see, e.g., Ref.\ \cite{ref:vogt2}.
Another major uncertainty which is difficult to quantify  stems from the form of the 
chosen ansatz (\ref{eq:htdef}) for the HT terms (see the discussion above). 
In addition we cannot take into account the correct $Q^2$ 
dependence of the HT terms 
since the relevant anomalous dimensions are not know (cf.\ footnote \ref{foot:htevol}).
Apart from studies of the scale dependence, see, e.g., Refs.\ \cite{ref:mv,ref:alekhin2},  
none of the other sources of theoretical uncertainties have been included 
in any of the combined extractions of HT terms and $\alpha_s(M_Z)$ from DIS data.
Since $C_{\mathrm{HT}}(x)$ and $\alpha_s(M_Z)$ are highly correlated
we therefore argue that errors on $\alpha_s(M_Z)$ from those types of analyses
are perhaps seriously underestimated and of limited use
(in particular when one takes into account also the obscured statistical significance of
$\chi^2$ in a QCD analysis \cite{ref:tung}).
To determine $\alpha_s(M_Z)$ from DIS it seems to be much safer to 
introduce first appropriate cuts on $Q^2$ and $W^2$ to remove the 
kinematical region which is contaminated by HT and then use the
obtained $\alpha_s(M_Z)$ value (or that from a global QCD analysis) to
determine the $C_{\mathrm{HT}}(x)$ afterwards by relaxing the cuts and
keeping $\alpha_s(M_Z)$ fixed.

Finally, let us compare our results with previous ones available 
in the literature. On the l.h.s.\ of Fig.\ 2 we also show the
HT coefficients obtained in the NLO fit in \cite{ref:mv}. 
Despite the differences in both analyses (see also footnote \ref{foot:mv}) 
reasonable agreement is achieved with our NLO results for $C_{\mathrm{HT}}$.
We cannot directly compare with the $C_{\mathrm{HT}}$ in
\cite{ref:alekhin,ref:alekhin2} since their ansatz differs from
(\ref{eq:htdef}), see also footnote \ref{foot:ansatz}, but the main
features are the same, in particular the change of sign in $C_{\mathrm{HT}}$
at around $x\simeq 0.45$. 
As already mentioned in the introduction, it was argued in 
\cite{ref:bodek2} that NNLO corrections reduce the need for 
HT contributions to $F_2$.
However, in \cite{ref:bodek2} only the NNLO coefficient functions
were used and the parton densities were taken from a NLO global
analysis without refitting them. 
Each $x$ bin was assigned a separate normalization
factor, called `floating factor', to substitute for the necessary refit 
of the parton densities \cite{ref:bodek2}, and the HT 
coefficients were taken from the IR renormalon model
\cite{ref:webber}. Nevertheless, we confirm their conclusions. 
In addition, SG seem to further reduce the need for HT at large $x$. 

\begin{figure}[tb]
\begin{center}
\epsfig{file=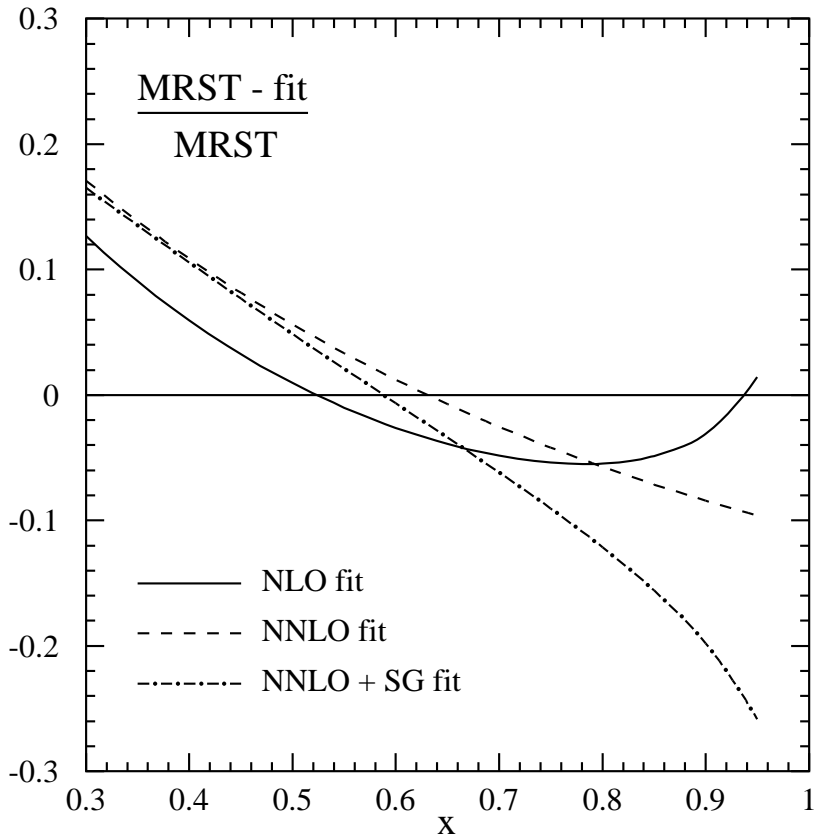,width=0.75\textwidth}
\caption{\sf Comparison of our fitted input distribution $f$
according to Eq.\ (\ref{eq:ansatz}) with the corresponding non-singlet 
(valence) combination of a recent MRST global analysis \cite{ref:mrst}.
Shown are the deviations from the MRST `central fit' at
the input scale $\mu_f = \mu_0 = 1\;\mathrm{GeV}$.}
\end{center}
\end{figure}
In Fig.\ 3 our fitted input distributions $f$ are compared with the
corresponding combination of parton densities from a recent
{\em global} NLO QCD analysis \cite{ref:mrst} which to a large extent excludes
the kinematical region where HT become operative. Over the entire
$x$ range included in our analysis, $0.35\le x\le 0.85$, our NLO
$f$ differs from the MRST result by less than $10\%$. Only at NNLO
with SG included, our $f$ turns out to be significantly smaller
than the MRST result as $x\rightarrow 1$. This gives an indication of the
importance of SG at large $x$. Since the resummations increase the 
coefficient functions smaller parton densities are needed in that region 
to describe the data.

To summarize, we have studied the impact of higher order
corrections on the extraction of HT contributions to the DIS
structure function $F_2$. The size of the required HT terms is diminishing when
NNLO corrections are taken into account as was recently argued in the literature. 
However, even at NNLO with SG  still a sizable, positive HT term is needed at
large $x$ while for $x\lesssim 0.45$ a small negative HT contribution is required.
SG resummations tend to further reduce the size of the HT
contributions and should be therefore included in future analysis
of upcoming data at large $x$, e.g., from `CEBAF @ 12 GeV'.
It was argued that simultaneous extractions of $\alpha_s(M_Z)$ and
HT terms in DIS may suffer from large theoretical uncertainties 
due to the strong model dependence and the unknown $Q^2$ evolution of the
HT contribution.

\section*{Acknowledgments}
We are grateful to A.\ Vogt for providing us with the {\sc Fortran}
routines of the results presented in Refs.\ \cite{ref:vogt,ref:vogtsg} 
and for valuable discussions. 
We thank S.I.\ Alekhin for useful discussions concerning 
Refs.\ \cite{ref:alekhin,ref:alekhin2} and the importance
of correlations of systematic errors.
The work of M.S.\ was supported in part by the National
Science Foundation grant no.\ PHY-9722101.
This work was supported by DFG and bmb+f.

%

\end{document}